\documentclass[letterpaper]{article}

\usepackage{natbib,alifeconf}  
\usepackage{hyperref}
\hypersetup{%
 setpagesize=false,%
 bookmarks=true,%
 bookmarksdepth=tocdepth,%
 bookmarksnumbered=true,%
 colorlinks=true,%
 hidelinks,  
 linkcolor = red,
 pdftitle={},%
 pdfsubject={},%
 pdfauthor={},%
 pdfkeywords={}}

\title{ALIFE 2023}

\title{Automata Quest: NCAs as a Video Game Life Mechanic}
\author{Hiroki Sato$^{1, 2, *}$, Tanner Lund$^{1, *}$, Takahide Yoshida$^{1}$, \and Atsushi Masumori$^{1, 2, *}$ \\
\mbox{}\\
$^1$Ikegami Lab, Department of General Systems Studies, University of Tokyo, Komaba, Tokyo, Japan \\
$^2$Alternative Machine Inc., Tokyo, Japan \\
 $^*$These two authors contributed equally to this work \\
hsato@sacral.c.u-tokyo.ac.jp} 

\begin{document}
\maketitle

\begin{abstract}
We study life over the course of video game history as represented by their mechanics. While there have been some variations depending on genre or "character type", we find that most games converge to a similar representation. We also examine the development of Conway's Game of Life (one of the first zero player games) and related automata that have developed over the years. With this history in mind, we investigate the viability of one popular form of automata, namely Neural Cellular Automata, as a way to more fully express life within video game settings and innovate new game mechanics or gameplay loops. Implementatin of this demo is available at \href{https://github.com/IkegLab/GNCA_invader/blob/master/README.md}{this URL}.

\end{abstract}

\section{Introduction}

The world of video games encompasses a vast array of immersive experiences, often involving the player engaging in battles against various enemy characters such as wyverns, zombies, robots, aliens, and humans. These characters, as well as those controlled by the player, are portrayed as alive entities within the game, and are treated as such until certain conditions are met. Yet, is that all there is to “being alive” according to video game mechanics?

In early video games such as Space Invaders, the invading characters’ life (i.e. alive to dead) representation is binary—they are either hit by the player's shots and destroyed or remain unscathed and fully functional. As computational power has increased and game design evolved, more nuanced representations of being alive were made possible. The most significant change was the introduction of hit points, popularized by games like Dungeons \& Dragons \citep{9780880387293}, to measure “how close” a given character is to their death state. However, despite advancements in game design and technology, the reliance on hit points as a primary metric for character vitality has persisted with only minor innovation throughout the history of video game development. Other paradigms and representations are few. This raises intriguing questions about the nature of life representation in gaming and whether alternative approaches can enhance player immersion and engagement. 

The Artificial Life (ALife) movement which aims to explores the ultimate question of ``what is life?'' through artificial creation of life-like phenomena has accumulated many theoretical and implementation frameworks of life in decades. In particular, the increase of computational power has enabled the development of many ALife frameworks and \textit{in-silico} implementations of various models of life. Here, we argue that these models are potentially useful to innovate representations and mechanisms for representing life in video games. 

In this paper, we aim to explore the current state of representation of life in terms of being alive to dead in video games, delving into its historical origins and examining notable examples. By analyzing the ways in which being alive is portrayed and quantified in different game genres and eras, we gain insights into the potential for novel approaches to character vitality. We then introduce self-organization, specifically Neutral Cellular Automata (NCA) \citep{mordvintsev2020growing} as a novel method of representing the life and strength of characters through controlled (designed) anti-damage training. This enables a representation of characters' life with regeneration of characters parts rather than numerical character parameters currently used. Ultimately, we hope to pave the way for advancements in game mechanics that more accurately capture the essence of being alive within virtual worlds, enhancing player immersion and contributing to the continued growth and evolution of the medium.

\section{Life as Represented by Game Mechanics}
The way game mechanics represent life is an example of "procedural rhetoric" and reveals the game's perspective on the life of digital beings \cite{doi:10.1162/dmal.9780262693646.117}. Games are understood to be liminal and therefore only need to be "real enough" to create a compelling experience for players, but nevertheless can tell us something about what digital life "is", allowing us to then reason about what it "could be" \citep{doi:10.1177/1555412014557542}.

We begin our study by examining the various ways that life is represented in game mechanics, including the role of hit points, boss and player character (PC) design, separate measures of life, status effects, the distinction between being knocked down and being "dead", and the difference between the life of the player and the life of what they control. While we acknowledge that the player's \textit{perception} during gameplay of life and its vibrancy/fragility/value can be affected by many additional factors such as sound design, camera angles, and narrative, as epitomized by the survival-horror genre \citep{315e6dd32071450baefd68546019f80c}, we have limited our current study to gameplay mechanics only. We do this in the pursuit of opportunities for mechanic innovation as well as what different mechanics say about digital life.


\subsection{Hit Points}
One of the primary mechanics in modern video games is that of Hit points (HP), which were popularized the tabletop game Dungeons and Dragons \citep{9780880387293} and later became a staple the industry. In earlier games, characters were either alive or dead (binary), but with the introduction of HP, the PC could take damage without necessarily dying. This system allows for more complex gameplay and combat, as characters can take multiple hits before succumbing to their injuries. This allows for more complex game states and gameplay decisions. For example, in the game World of Warcraft \citep{wow}, players have a health bar that shows their current hit points, and they can use healing spells or items to restore their HP and continue fighting. Managing HP is a core aspect of gameplay, and remains so even in player-vs-player (PvP) contexts. 

\subsection{Additional Measures of Life}
There are other measures, separate from HP, that impact a character's ability to perform actions, which is another way to measure life. In Role-Playing Games (RPGs) such as the aforementioned World of Warcraft, magic points (MP) can limit the use of special abilities or spells, while money can restrict the purchase of items or equipment. The restriction of item purchase to certain locations/times and of spells to certain characters or character levels adds further complexity.

Morale and sanity measures can also impact the ability to perform certain actions, such as causing a character to become disheartened or go insane. For instance, in the game Darkest Dungeon \citep{darkestdungeon}, characters have a stress meter that changes over time and causes them to become stressed or irrational, leading to impaired combat performance. In the game Amnesia \citep{amnesia}, sanity is depleted by witnessing disturbing events or spending time in the dark. When the sanity meter reaches zero, the player becomes unable to control the character effectively, making it difficult to progress through the game. To restore sanity, the player must find light sources or consume items that restore sanity.

\subsection{Status Effects}
Status effects are another way that agent abilities are affected, common to RPGs but not exclusive thereto. These can include anything from poison or burn damage to debuffs that hinder a character's abilities. Different games have different status effects, each with its own set of rules and consequences. For example, in the game Pokemon \citep{pokemon}, status effects such as paralysis or poison can hinder a Pokemon's ability to fight, while other status effects such as burn can cause damage over time. Once again, the remedy for such effects is often an item, a spell, or the passage of time.


\subsection{Unique Bosses and PCs}
Bosses and PCs have more variety in how their life is represented, with many games including boss phases, injury, changing attack patterns, and limits on abilities. Bosses often have multiple forms or phases, each with its own set of attacks and strategies required to defeat them. This design encourages players to adapt and change their tactics, making the game more challenging and engaging. An example of this can be seen in the game Dark Souls \citep{darksouls}, where bosses have multiple phases that become progressively harder to defeat, with different attack patterns and strategies required for each phase.

In some games, injury can also impact a boss or PC's abilities or skills. For example, in the game Fallout 4 \citep{fallout}, if a PC's limb is injured, their ability to use that limb may be impaired, affecting their ability to aim or perform other actions. In the game Red Dead Redemption 2, a PC's stamina may be reduced if they sustain injuries from falls or attacks, making it harder for them to perform physical tasks or escape danger. Also, injury is sometimes a required step to defeat enemies and progress in games. In the game series Monster Hunter \citep{monsterhunter}, injury of target monsters is implicitly designed as the trigger of boss phase shifts by debuffing monsters' abilities related to the injured body part permanently (the opposite approach of Dark Souls) and changes the reward for successful hunts. 

\subsection{The Effects of Death on the Game World}
Non-Player Characters (NPCs) in video games are often expendable and can be killed without any major consequences to the story or gameplay. This is particularly true for NPCs that are not central to the game's plot or mission objectives, such as bosses. For example, in the game Skyrim \citep{skyrim}, players can kill most NPCs in the game without affecting the main questline or their ability to complete side quests. While some NPCs may have certain roles or items that are needed for quests or story progression, there are often alternative ways to obtain these objectives without the NPC. Most NPCs do not have phases, and only the occasional NPC will use items or spells to heal themselves. Their deaths are immaterial, and under some conditions new characters just like them may spawn in their place, rendering those deaths even less meaningful from the player’s perspective.

On the other hand, the death of a PC can have a significant impact on gameplay itself. The player may have invested time and resources into building and developing their character, and losing that progress can be a major setback. As a result, many games have implemented mechanics to prevent or mitigate character death for PCs. For example, games like Minecraft \citep{minecraft} use a "death penalty" mechanic where the player may drop their items upon death, but can still respawn and continue playing. Managing the risks of death by setting spawn points and storing items in chests is part of playing the game.

Other games like Dark Souls are much less forgiving, and use a "permadeath" mechanic where the player loses all progress upon death and must start over from the beginning or the nearest checkpoint. In this latter example, the game world does not continue on after the player character dies. Repeated attempts at the same section of the game is an expected part of the experience, and game difficulty is modulated by the frequency of checkpoint placements.

In some cases, the death of a PC can also have narrative consequences. A horrible end to the Dark Souls game world is presumed but not shown when the PC dies therein. As a contrasting, more nuanced example, in the game Mass Effect 2 \citep{masseffect} the death of certain party members can affect the game's story and the player's ability to complete certain objectives. This can lead to players feeling emotionally invested in their characters and more motivated to avoid their death - even though the game can continue on without them.

\subsection{"Knocked Down" vs "Dead"}
The distinction between being knocked down and being "dead" is also an important aspect of life representation in games, as some games allow players to revive fallen allies, while others do not. This mechanic encourages teamwork and cooperation, as players must work together to revive their fallen comrades, whether controlled by the same player or by others. In the game Left 4 Dead 2 \citep{left4dead}, players can revive their fallen teammates within a limited time frame, but if they fail to do so, the player is considered dead and cannot be revived until the next level.

\subsection{Multiple Lives}
The question of whether multiple lives count as one creature or a different one is also relevant in some games. For example, in games like Super Mario Bros. \citep{smb}, the player can gain extra lives, but these are treated as separate entities rather than being part of the same character. This design choice has implications for how the game is played and how players approach challenges, but this paradigm is more similar to Reinforcement Learning's multiple runs \citep{sutton2018reinforcement} than it is to a representation of life.

\subsection{Agents vs Units}
In strategy games like Warhammer 40k: Dawn of War \citep{dow}, players control multiple units that act as a single entity. These units often consist of several individual agents or characters, each with their own HP, abilities, and strengths. However, in terms of "life", the unit as a whole is typically considered as a single entity. The unit's overall HP represents the combined health of all the individual characters within it, and when the unit's HP reaches zero, the entire unit is destroyed, regardless of how many individual characters were still alive. Similarly, healing the unit may bring back individual characters that had disappeared.

\subsection{The Life of the Player vs. the Life of What They Control}
Finally, it should be noted that the life of the player and the life of what they control are not the same in some games. In real-time strategy (RTS) games, for example, the player is not eliminated until all their units or buildings are destroyed. This means that even if the player loses all of their units, they can still continue to play and try to rebuild their army. Multiple lives are another representation of this concept, and in fact the earliest one, allowing the player to try again even though their character has died, up to a limit. In multiplayer games, a player may continually re-spawn until some condition is met, such as a timer to end the match.

\section{Life as Represented by ALife Automata}
In this section, we consider the concept of being alive in ALife side with weight to cellualr automata and self-organization and discuss potential use of NCA for realization and expansion of mechanic and visual represetaiton of video game characters' well-being, ability and existence. 

\subsection{From The Game of Life to NCA}
Ever since the invention of cellular automata (CA) by Stanislaw Ulam and John von Neumann and development of Game of Life (GoL) by John Conway \citep{Wolfram2002}, CA has attracted researchers and programmers' interests. From the perspective of games, GoL is considered a zero-player game \citep{Bjork2012-ba}, which is a small but enduring genre of games that challenge traditional conventions and mechanics. This gaming heritage perhaps makes NCAs a good candidate for future innovations.

NCAs are a type of cellular automata capable of learning their update rules via artificial neural networks. Among NCAs, the Growing Neural Cellular Automata (GNCA), first introduced by \citet{mordvintsev2020growing}, are notable in their ability to grow from a seed cell to certain patterns and ``re-grow'' destroyed portions of themselves after disruption. This behavior shows potential as a gameplay mechanic and representation of the maintenance of life. Rather than representing an enemy's well-being through numerical parameters like hit points or signifying a transition from fully capable to dead through injury, phase shifts or a death animation, NCAs potentially realize a more mechanistic and visual representation of characters' well-being, capability and existence by their self-organization. 

Self-organization has been a widely discussed topic in ALife as a potential mechanism of life, and it is one of main questions that NCA studies consider. Self-organization can develop and maintain the patterns, forms and functions of systems that would be destroyed without it. The mechanics of life in the game characters described above are obviously not result of self-organization, and characters are alive because their parameters are over a deactivation threshold which is unrelated to mechanics of their forms and functions. In reality, creatures are "supposed" to have no such built-in parameters and they are alive and functioning \textit{because} they can maintain their system as a whole including interrelated physical, cognitive and behavioral properties. Thus, building a creature without explicit "alive/dead" parameters brings something closer to the reality of being alive to games. 

However, reality is not optimal to video games in many cases, which are liminal and only approximate the real world. For instance, if player characters were as fragile as real humans, variation of gameplay experience would be very limited. Thus, simply introducing self-organization or certain mechanics about living-as-system to video games is not suitable in itself for consideration in game design. 

Here, the strength of NCA is its "trainability" which makes the NCA easier to design. In other words, game designers can train an NCA creature to be intentionally weak or robust against certain interventions to its system or structure. For instance, NCA creatures which are well trained at body but not at head can easily be beheaded. Also, it might be possible to strongly train the dermal cells (edges or surfaces) of a creature such that injuries on the hard shell skin let attackers to access to not well-trained, thus weak, part of the body. Furthermore, the phenomena of overgrowth observed on NCA can cause emergent phase shifts of the creature due to damage. Such NCAs' potential capability to be designed is considerable for game design. 

Note that even though NCA creatures are not alive or dead by explicit parameters, currently developed NCAs still have parameters on cells such as their growth speed and size. These parameters can also affect the robustness of creatures by increasing their heal speed. However, these parameters are part of creature mechanics as system unlike universal health point, and these parameters can be replaced by self-organization as well in future. 

\subsection{Scalability and Stability Limits}
While NCAs have shown promising results in generating visually appealing and coherent patterns, they does have certain spatial scalability limitations, given their dependence on the input image size. The behavior and output of NCA models are typically constrained by the size of the input image they are trained on. This means that if a model is trained on a specific image size, such as a 40x40 pixel image of a lizard emoji, it may have difficulty generating larger images or textures beyond that size.

This limitation is baked into early example of NCA models which designed to work within the dimensions of their input image. Therefore, when attempting to generate larger images, a model would require modifications to its architecture or training process to handle the increased complexity and transformation of spatial dimensions. This is challenging, as increasing the image size exponentially expands state space that the model needs to learn and operate within. Less important to proofs of concept but still notable for practical applications: it would likely be more computationally expensive. 

Furthermore, it is well understood and was demonstrated early on that certain amounts or forms of damage or deletion to an NCA structure can cause it to fail - to no longer be able to reconstruct itself. The resulting behavior ranges from developing a stable but unrecognizable form to a continually growing mass of pixels, or even a split into multiple instances of the same image. Such behavior possibly affected by things such as the shape and complexity (simplicity) of the image represented by the NCA. However, there has been no analytically well-established knowledge about stability of GNCAs. 

To overcome these limitations, researchers may pursue various approaches. One such approach is to develop hierarchical or multi-scale NCA models that can capture patterns and structures at different levels of detail, allowing for the generation of larger images by combining the outputs of multiple scales. Careful selection of the shapes of NCAs used and the ways in which they may be damaged can also mitigate regeneration errors, though this is not a wholly satisfying solution. 

Another, more technical solution would be use of anti-alias techniques on pixels. Instead of increasing image sizes, smoothing pixels can improve quality of rendered graphics. In addition, mapping textures on cells and using cells as position and state specifier can increase/decrease training image complexity. Also, use of GNCA as growing and regenerating state map may have potential for game development rather than just a solution to issues described above. 

\subsection{Movement Limits}
NCAs (and cellular automata more generally) are limited in their ability to move about their environment, somewhat limiting their utility as agents in a video game setting. NCAs or CAs in general are enclosed in cellular units, and “movement” is realized by spatial flow like shift of cell states, like gliders and moving lizards, or deformation of units, like voxel robots \citep{voxel}. The former is more analogus to movement by growth, such as is observed in plants \citep{tropism}, but on a more rapid and less resource-constrained timeline. The latter is close approach to muscular manipulation of skeletal tissues. Most video games focus on the mechanism of movement through bones which themselves are not deformed. 

This provides a practical challenge for introducing new movement patterns and paradigms as they go against established industrial norms and tooling to facilitate bone-based movement. Designing the motion of a character with deformation of units - such as one might observe in a voxel robot - \citep{Horibe2021-bk} requires significant consideration and development effort. Still, application of deforming muscle-like voxel robots with regenerating NCAs has big potential to game character implementation, even though we have to accomplish higher computational load of voxel characters in contrast to polygons. At the time of writing, NCAs cannot easily represent mobile agents, which limits their utility. 

\begin{figure}[h]
    \centering
    \includegraphics[width=1.1\linewidth]{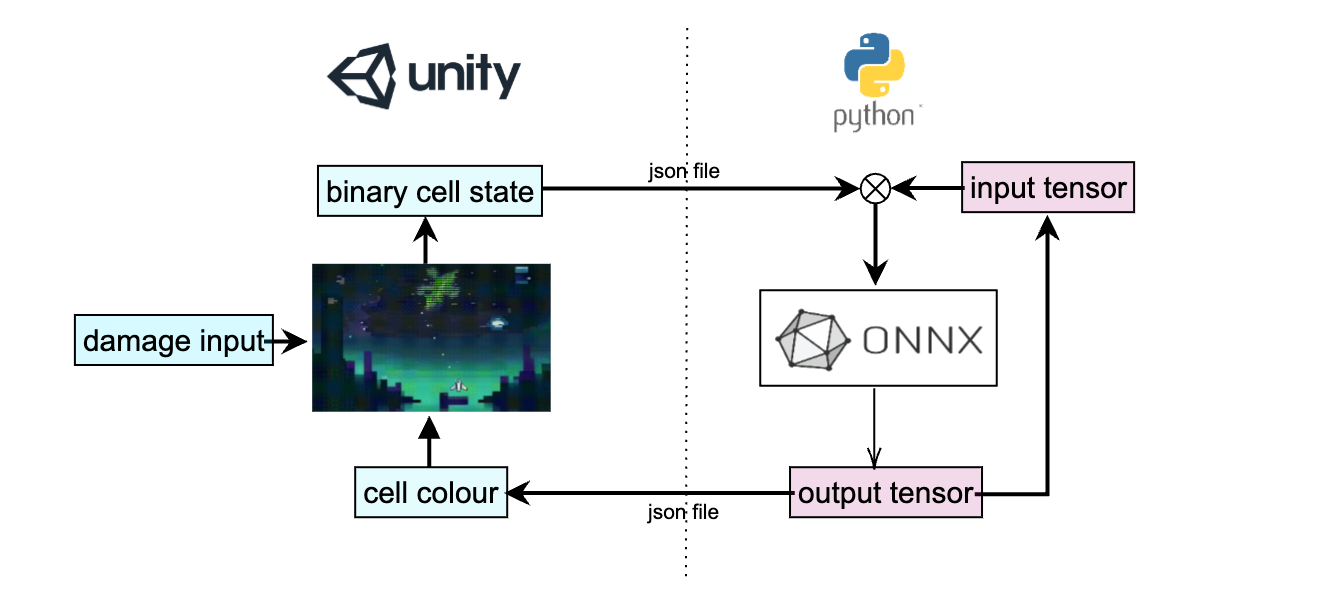}
    \caption{A schematic diagram of the implemented game system using Python and Unity. With regards to the Python code, we used a GNCA model with 16 channels per cell trained on PyTorch and converted to ONNX. The Python program receives an input of an array of binary cell state ($0=\:$dead, $1=\:$alive) on the game engine and applies it to the cell 16-channel states. The GNCA model infers the cells and outputs updated cell states. On the game engine (Unity) side, the engine reads the RGB$\alpha$ channels of the updated cell state and depicts alive cells as cubes with a color specified by the input. The cell cubes are broken on collision with bullets shot from a player-controlled space fighter. }
    \label{fig:architecture}
\end{figure}

\begin{figure}[h]
    \centering
    \includegraphics[height=.2\textheight]{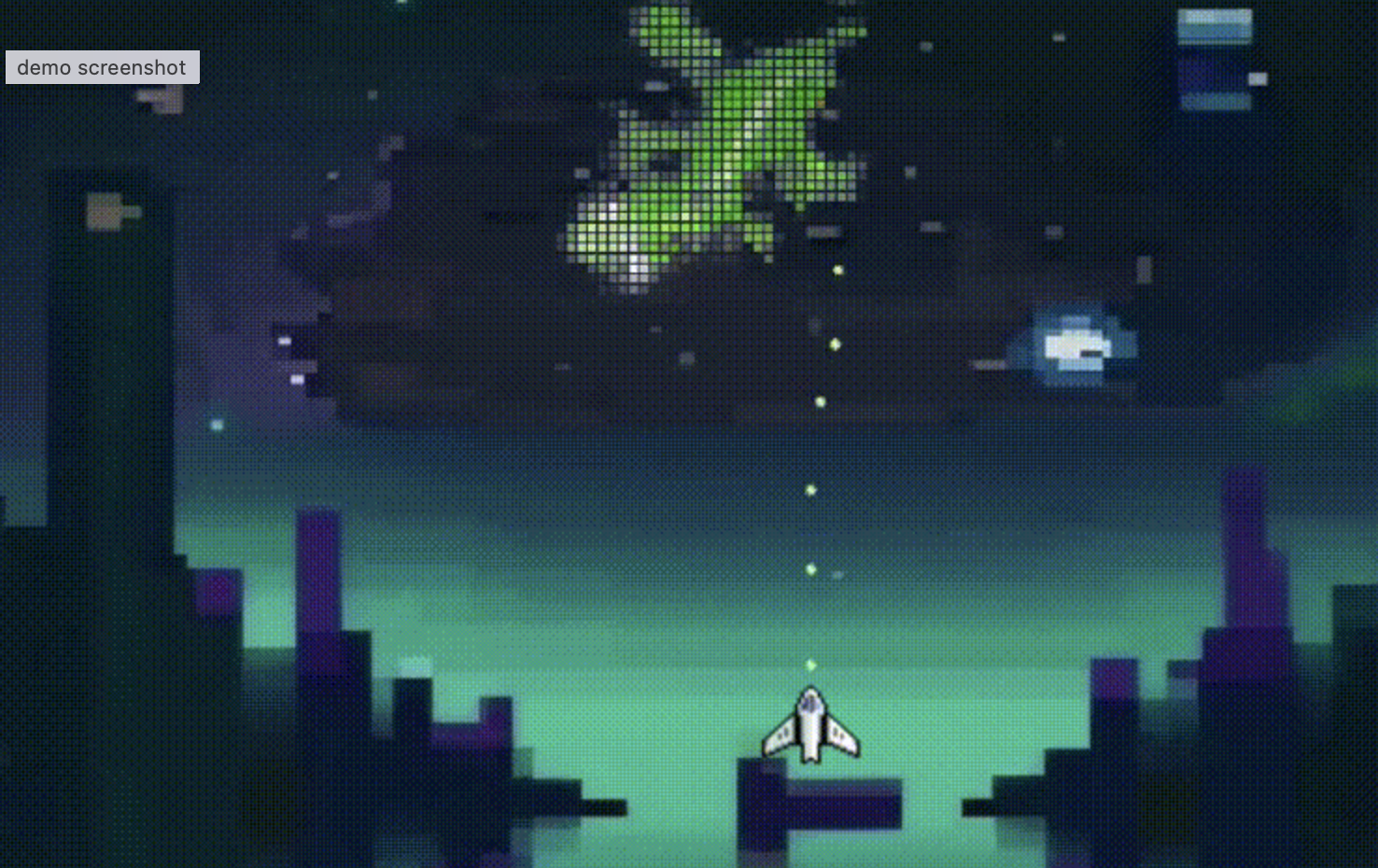}
    \caption{The most common example of an NCA, the lizard emoji, imported into a space-invaders style game implemented with Python and Unity. The player attacks and destroys the lizard, but the physical state of lizard begins to recover.}
    \label{fig:space_invaders}
\end{figure}

\section{Implementing Neural Cellular Automata In-Engine}

Introducing NCA to game engines requires running neural network models alongside game engine functions. Use of Open Neural Network Exchange (ONNX) is the most accessible option to use neural network models on game engines at present. Both of major game engines Unity and Unreal Engine (experimentally) support ONNX format. ONNX is also supported on Web systems like Node.js and it is possible to include it in game engines designed for Web usage like Babylon.js.

Here, as a preliminary implementation of GNCA to video games, we used Python for running neural network models and Unity for running the game (see \ref{fig:architecture}). Although there are performance, ease of implementation with this structure was prioritized so that communication between ALife, in which Python is main tool, and game developer community can be enhanced. Figure \ref{fig:space_invaders} is a screenshot of the implemented game. This implemantation is publicly available, you can try it (see Appendix).

The implemented game here is inspired by the classic Space Invaders from 1978. In the original game, space fighters shoot aliens. We replaced aliens with a GNCA lizard and player is aimed to defeat the lizard by eliminating all its cells. 

\subsection{Limitations}
During gameplay, players experience the regeneration of the NCA lizard not only toward the original lizard shape but also sometimes toward deformed shapes. There is also an intrinsic weak point in the NCA model used. If the center of the body is continuously disrupted, the body tends to deform and loses its lizard shape. Although the shape of the lizard is not critical issue for player in this particular game per se, the size can become a problem. For example, if the lizard body overgrows to cover the entire game field the player will have no way to avoid lethal contact to the creature. These can be considered shortcomings, but could, as discussed above, also be deliberatly incorporated into gameplay design.

Game difficulty is mostly controlled by the NCA generation speed and how well-trained (stable) the model is. If the creature regenerates too fast, the player has lower odds of eliminating all cells because they come back right after elimination. How well-trained the NCAs' models is is critical to properly regenerate the shape. A poorly trained model results in a lizard that is unstable and can disappear even without any attacks from player or overgrow without any intervention. Parameter tuning is of course necessary.

As mentioned above, the movement of the NCA creature needs further development and tuning. NCA in our game is also immobile. In the original Space Invaders, aliens moves horizontally back and forth or in an interesting pattern and it made the game experience richer. Currently possible options are letting the cell grid as a whole move which would make the game more like the original or attaching bones for movement to the cell grid to move it around as modern games do.

\section{Discussion}
A logical next step for this work would be to add regenerative NCA features to player characters, enabling different gameplay patterns and decisions based on regeneration speed, damage location, the resources needed to "regenerate", etc. Coupled with other mechanics (damage types, the risk of damage being too significant to regrow a certain limb properly, etc.), this could diversify potential gameplay loops

It should also be possible to couple NCAs with the existing game industry paradigm of mobile characters. NCAs can be used as skins (meshes) of "normal" 3D characters and shading targets. An NCA can be used as a representation of the state map of each polygonal mesh, like shape of marks on the character's surface, visible or invisible (perhaps demonstrating that the character still not injured/destroyed) and toughness of the polygonal body of characters.  

Applying insights from ALife research, including NCA, we can in principle create a game that never ends. Typically, characters in traditional games correspond one-to-one with input and output, behaving deterministically in response to player actions. However, by using ALife-like autonomous agents, we can achieve more diverse and nondeterministic behaviors. For example, in competitive online games, even if the game system is consistent, the agents present in the game are operated by humans and behave autonomously, keeping us from getting bored.

Furthermore, in games where the termination condition is not binary, like our shooting game using NCA, it is difficult to create a unique winning strategy. The lizards of the NCA have weaknesses, but in some cases attacking them can result in further proliferation or changes in shape, making prediction difficult and adding a different type of dynamism than the semi-randomness of game genres like rogue-likes. Thus, through the autonomy and robustness of life that the field of artificial life has continued to research, we can create endless games that don’t bore players and do not have a unique winning strategy.

NCAs have potential not only for representations of one character, but perhaps also many. NCAs could represent a large swarm of semi-independent enemies. Whether this were an action game, a zombie game, or some other type with many enemies, a swarm that can regenerate at a certain pace would make for an interesting challenge or interaction. Indeed, destructible environments could also benefit from such behavior: the walls of a an "organic" building, for example. Opportunities abound.

\section{Acknowledgements}

This work was partially supported by JSPS KAKENHI; Grant Number 21K17822.

\section{Appendix}
The source code of the implemented  invader game with NCA is available. Please download and play the game from the following link: \url{https://github.com/IkegLab/GNCA_invader/blob/master/README.md}

\footnotesize
\bibliographystyle{apalike}
\bibliography{reference} 

\end{document}